\documentclass[10pt,conference]{IEEEtran}
\usepackage{cite}
\usepackage[tight,footnotesize]{subfigure}
\usepackage{graphicx}
\usepackage[cmex10]{amsmath}
\usepackage{amssymb}
\usepackage{mathtools}
\usepackage{multirow}
\usepackage{notoccite}
\usepackage{color} 
\usepackage{epsfig}
\usepackage{latexsym}
\usepackage{subfigure}
\usepackage{pstricks}
\usepackage{pdfpages}
\usepackage{epstopdf}
\DeclareGraphicsExtensions{.eps}
\usepackage{algorithmic}
\usepackage{algorithm}
\usepackage{multirow}
\usepackage{graphicx}
\usepackage{xurl}

\begin{document}
	\title{Integrated Access and Backhaul via LEO Satellites with Inter-Satellite Links}
	\author{\IEEEauthorblockN{Zaid Abdullah$^\dagger$, Eva Lagunas$^\dagger$, Steven Kisseleff$^\star$,   Frank Zeppenfeldt$^\ddagger$, and Symeon Chatzinotas$^\dagger$ \\
			\\ $^\dagger$ Interdisciplinary Centre for Security, Reliability and Trust, University of Luxembourg, Luxembourg \\
   Emails: \{zaid.abdullah, eva.lagunas, symeon.chatzinotas\}@uni.lu\\ 
   $^\star$ Fraunhofer Institute for Integrated Circuits IIS, Erlangen, Germany. Email: steven.kisseleff@iis.fraunhofer.de\\
			$^\ddagger$ European Space Agency (ESA), Noordwijk ZH, The Netherlands. Email: frank.zeppenfeldt@esa.int}}
	
\maketitle
\begin{abstract}
The third generation partnership project (3GPP) has recently defined two frequency bands for direct access with satellites, which is a concrete step toward realizing the anticipated space-air-ground integrated networks. In addition, given the rapid increase in the numbers of satellites orbiting the Earth and emerging satellites applications, non-terrestrial networks (NTNs) might soon need to operate with integrated access and backhaul (IAB), which has been standardized for terrestrial networks to enable low-cost, flexible and scalable network densification. Therefore, this work investigates the performance of satellite IAB, where the same spectrum resources at a low earth orbit (LEO) satellite are utilized to provide access to a handheld user (UE) and backhaul via inter-satellite links. The UE is assumed to operate with frequency division duplex (FDD) as specified by the 3GPP, while both FDD and time division duplex (TDD) are investigated for backhauling. Our analysis demonstrate that the interference between access and backhaul links can significantly affect the performance under TDD backhauling, especially when the access link comes with a high quality-of-service demands.

\end{abstract}

\section{Introduction}
\subsection{Definition and Standardization}
Integrated access and backhaul (IAB) allows spectrum sharing between access links to end users (UEs) and backhaul links between the radio access network (RAN) nodes. The third generation partnership project (3GPP) has standardized the IAB as part of the fifth-generation new radio (5G-NR) to enable cost- and time-efficient network densification~\cite{report}, leading to a better quality-of-service (QoS) with reduced latency~\cite{zhang2021survey, madapatha2020integrated}. The RAN nodes that provide direct access to UEs and backhaul to other (parent or child) RAN nodes are called IAB nodes, while the gNodeB (gNB) that connects directly to the 5G core network via a non-IAB link (such as optical fiber) is called the IAB donor. 

Two types of IAB operations have been defined by the 3GPP in 5G-NR: In-band (IB) and out-of-band (OB). For the OB-IAB, the access and backhaul links are assigned to different frequency bands and thus, they do not interfer with each other. In contrast, the IB-IAB should have at least a partial frequency overlap between the two links (i.e. access and backhaul), and therefore, one should account for the resultant interference to ensure that the required QoS constraints are satisfied. Regarding the operational frequencies for IAB, different bands in the first and second frequency ranges (FR1: below $7.125$ GHz and FR2: $24.25 - 52.6$ GHz) with time division duplex (TDD) have been adopted by the 3GPP specifications, which can be found in release 17~\cite{Rel17}. 
\subsection{From Terrestrial to Non-Terrestrial Networks (NTNs)} 
Thus far, the IAB operation has been part of the 3GPP standardization only for terrestrial networks. Nonetheless, for direct access to handheld UEs through NTNs, two FR1 frequency bands ($n256$ and $n255$) with frequency division duplex (FDD) have been allocated by the 3GPP in release 18~\cite{Rel18}, which is a key step toward the anticipated integration between the terrestrial and non-terrestrial domains. In this context, a recent preliminary investigation on utilizing the $n256$ and $n255$ bands for 5G access through NTNs was carried out in~\cite{pastukh2023challenges}. The results showed that these two bands might lead to significant interference on various existing satellite infrastructure, prompting the exploration of  additional or alternative frequencies for direct access through NTNs in 5G and beyond.  

Whether or not the currently adopted frequency bands will be modified in the future to avoid such interference, one can be certain that NTNs will play a pivotal role in the near future~\cite{kodheli2020satellite}, not only in providing connectivity to rural or isolated areas, but also to hotspot areas where terrestrial networks on their own cannot cope with the ongoing data traffic. From all the above, it becomes quite clear that, similar to the evolution of terrestrial networks, NTNs might also soon need to adopt an IAB operation that is compatible with terrestrial network operators, to provide global connectivity with enhanced spectrum utilization. This was the motivation behind our preliminary work in~\cite{PIMRC}, where we analyzed the main challenges associated with an NTN-based IAB operation. In addition, a case study was provided where a low earth orbit (LEO) satellite provides access to a handheld UE and backhaul to a terrestrial base station (BS) with shared spectrum.  
\subsection{Contribution}
In this work, we investigate a different aspect of satellite IAB operation, where a LEO satellite provides direct access to a handheld UE, and backhaul to a second LEO satellite with the same spectrum resources (i.e. IB-IAB) over the S-band in FR1. For the access link, we adopt the FDD transmission according to the 3GPP specification~\cite{Rel18}, while both TDD and FDD are explored for backhauling through inter-satellite links (ISLs). We study the effects of number of LEO satellites per orbital plane, the altitude of LEO satellites, the required QoS at UE, and the ISL transmission mode on the network throughput under the IAB operation. 

It should be mentioned that in our previous work on satellite IAB in~\cite{PIMRC}, the backhauling was between a LEO satellite and a terrestrial BS, which is substantially different from the system and analysis presented here. To the best of our knowledge, the work presented here and our previous work in~\cite{PIMRC} are the only available works on satellite IAB so far. 

The rest of this paper is organized as follows. Section~\ref{sec: system_model} presents the system and channel models. Section~\ref{sec: FDD} and Section~\ref{sec: TDD} deal with the received signals and resource optimization under FDD and TDD backhauling, respectively. Numerical results are presented and discussed in Section~\ref{sec: Results}. Finally, conclusions are drawn in Section~\ref{sec: Conclusions}.

\section{System and Channel Models}\label{sec: system_model}
\subsection{System Model}
\begin{figure}[t]
 \centering
       \includegraphics[width=6cm,height=5.0cm]{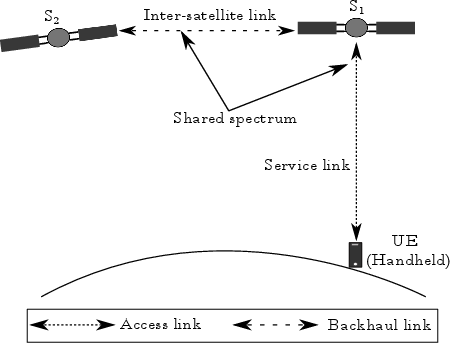}
        \caption{Satellite IAB with ISL-UE shared spectrum.}
        \label{Arch2}
\end{figure}

We consider a system with two LEO satellites orbiting the same plane, denoted by $S_1$ and $S_2$, with IAB capabilities (i.e. $S_1$ and $S_2$ are IAB nodes). Both $S_1$ and $S_2$ are assumed to communicate with the satellite gateway (or IAB donor) via an ideal feeder link with a dedicated frequency band. Our main focus will be on the IAB operation at $S_1$, which utilizes the same spectrum resources for data backhauling through the ISL with $S_2$, and also for data access to the UE via the service link (see Fig. \ref{Arch2}). We assume that the UE always operates with FDD mode to communicate with $S_1$ according to the 3GPP standardization on UE access with LEO satellites~\cite{Rel18}. On the other hand, both TDD and FDD transmission modes will be investigated for data backhauling through the ISL. 

The ground-based handheld UE is assumed to be located below $S_1$, as indicated in Fig. \ref{Arch2}. Therefore, and due to the fact that the terrestrial UE and $S_2$ are widely separated in space, the two are served by different beams (and possibly by different antenna sets) at $S_1$ that are almost orthogonal to each other. As a result of the access and backhaul beams orthogonality, each of the two nodes (UE and $S_2$) can be assigned the full bandwidth when served by $S_1$ without causing any notable inter-beam interference, even when adopting an IB-IAB operation with shared spectrum. On the other hand, the signal transmitted from $S_2$ to $S_1$ can potentially cause interference at the UE through undesired side-lobes. Similarly, the transmitted signal from the UE (which is intended for $S_1$) can also affect the received signal at $S_2$ due to the omni-directional transmission property at the handheld UE. Although, the interference at $S_2$ will have a small effect on the performance for the considered single UE case, given the restriction on the transmit power level at the handheld device.\footnote{According to the 3GPP release 18, the maximum UE transmit power within the channel bandwidth of a 5G-NR carrier cannot be higher than $23$ dBm, which is almost equivalent to $0.2$ Watts~\cite{Rel18}.}$^,$\footnote{The effect of the transmission of a large number of UEs on satellites under IAB operation will be the topic of a future research.} The interference between the UE and $S_2$ appears when the TDD is utilized for backhauling. Further details in this regard will be provided in Section~\ref{sec: TDD}.
\subsection{Channel Model and Channels Gain}
All channels are assumed to be line-of-sight (LoS), and the free-space path loss (PL) model is utilized to characterize their gain. 
\subsubsection{UE channel gain}
The channel gain between the UE and $j$th satellite ($j\in\{1,2\}$) can be given as:
\begin{equation}
\small 
    \beta_{\text{UE}j} = \frac{G_{S_j}G_{UE}}{(4\pi d_j f_c/c)^2},
\end{equation}where $f_c$ is the carrier frequency, $c$ is the speed of light, $G_{S_j}$ is the antenna gain of $S_j$, $G_{UE}$ is the UE antenna gain, and $d_j$ is the distance between the UE and $S_j$. Given the fact that $S_1$ is assumed to be located precisely above the UE, $d_1$ is nothing but the satellite's altitude, which we denote by $l_s$, while $d_2$ can be evaluated using a simple trigonometric formula as shown in the Appendix at the end of this paper. 
\subsubsection{ISL path-loss and channel gain}
The free-space PL for the ISL ($\mathrm{PL}_{I}$) can be expressed as~\cite{leyva2021inter}:
\begin{equation}
\small 
\mathrm{PL}_{I} = \begin{cases}
 \left(\frac{4\pi d_I f_c}{c}\right)^2, \hspace{2cm} \text{if}\  d_I \le d_{I_{\text{max}}} \\ \infty, \hspace{3.2cm} \text{otherwise}
\end{cases}
\end{equation}where $d_I$ is the ISL distance between $S_1$ and $S_2$, and $d_{I_{\text{max}}}$ is the maximum slant range between $S_1$ and $S_2$ to maintain an LoS connection (details are provided below). It follows that the corresponding ISL channel gain is:
 \begin{equation}
     \small 
     \beta_I = \frac{G_{S_1} G_{S_2}}{\mathrm{PL}_I}.
 \end{equation} 
\subsubsection{Slant range and number of satellites per plane} For two satellite nodes orbiting the same plane (and thus have the same altitudes), the maximum slant range can be evaluated according to the following formula~\cite{leyva2021inter}:
\begin{equation}\label{max_slant}
\small 
d_{I_{\text{max}}} = 2\sqrt{l_{\text{s}}(l_{\text{s}} + 2R_{E})},
\end{equation}where $R_E$ is the Earth's radius. In general, the slant range between two neighbouring satellites in the same orbital plane depends on the satellites' altitude as well as the number of satellites in that plane. Assuming evenly distributed satellites, the slant range between two neighbouring satellites is~\cite{soret2019inter}:
\begin{equation}\label{slant}
\small 
d_I = 2(R_E + l_s)\sin{(\pi/N_p)}
\end{equation}with $N_p$ being the number of satellites in the plane. Therefore, from (\ref{max_slant}) and (\ref{slant}), the required minimum number of satellites ($N_{p_\text{min}}$) in the plane to maintain an LoS connection is:
 \begin{equation}\label{N_min}
 \small 
\left \lceil N_{p_\text{min}} \right \rceil = \frac{\pi}{\sin^{-1}\left(\frac{d_{I_{\text{max}}}}{2\big(R_E + l_s\big)}\right)},
 \end{equation}where $\left \lceil x \right \rceil$ is the ceiling function defined as the smallest integer that is not smaller than $x$. 
 
 In the following two sections, we will elaborate on the received signals and signal-to-noise ratios (SNRs) at both the UE and $S_2$, and formulate the corresponding achievable network throughput for two different scenarios. The first scenario assumes FDD-based transmission for the ISL, while the second scenario deals with the case where TDD is utilized for backhauling between $S_1$ and $S_2$.
\section{Scenario 1: FDD for ISL} \label{sec: FDD}
In this scenario, both satellites operate according to the FDD mode (i.e. FDD is utilized for both access to the UE and backhaul through the ISL). Our focus here will be on the data transmission from $S_1$. It is worth mentioning that in this scenario there is no interference between $S_2$ and the UE. The reason is that the UE and $S_2$ operate in FDD, and they can adopt the same uplink frequency to transmit to $S_1$, and also the same downlink frequency to receive from $S_1$. Therefore, the transmission from the UE ($S_2$) to $S_1$ will not cause any interference at $S_2$ (UE).
\subsection{Received Signals and SNRs}
The received signals at the UE and $S_2$ can be expressed, respectively, as follows:
\begin{subequations}
\begin{equation}\label{yue}\small 
y_{\text{UE}}^{\mathrm F} = \sqrt{P_{A}} h_{\text{UE}1} x_{\text{UE}} + z_{\text{UE}},
\end{equation}
\begin{equation}\label{yse}\small 
y_{S_2}^{\mathrm F} = \sqrt{P_{I_{1}}} h_{I} x_{I_2} + z_{S_2},
\end{equation}
\end{subequations}where the superscripts in $y_{\text{UE}}^{\mathrm F}$ and $y_{S_2}^{\mathrm F}$ indicate that FDD transmission is adopted for the ISL, $P_{A}$ and $P_{I_{1}}$ are, respectively, the allocated powers at $S_1$ for the UE and $S_2$, $h_{\text{UE}1}$ is the channel coefficient between $S_1$ and the UE, while $h_{I}$ is the ISL channel coefficient.
Also, $x_{\text{UE}}$ and $x_{I_2}$ are the information symbols intended for the UE and $S_2$, respectively, satisfying $\mathbb{E}\{|x_{\text{UE}}|^2\} = \mathbb{E}\{|x_{I_2}|^2\} = 1$, while $z_{\text{UE}}$ and $z_{S_2}$ account for the additive white Gaussian noise (AWGN) at the UE and $S_2$, respectively, with noise power densities (PSDs) of $N_0$ per Hz.

Then, the received SNRs at the UE and $S_2$ are, respectively, given as:
\begin{subequations}
    \begin{equation} \small 
        \gamma_{\text{UE}}^{\mathrm F} = \frac{P_{A}}{N_0W_F} \beta_{\text{UE1}},
    \end{equation}
    \begin{equation} \small 
        \gamma_{S_2}^{\mathrm F} = \frac{P_{I_{1}}}{N_0W_F} \beta_I,
    \end{equation}
\end{subequations}where $W_F$ reflects the amount of bandwidth assigned for the FDD transmission in one direction. 
\subsection{Achievable Rates and Power Control}
The total throughput of the network (in bits/sec) is:
{\small \begin{align}
\mathcal R^{\mathrm F} = & \mathcal R_{A}^{\mathrm F} + \mathcal R_{\text{ISL}}^{\mathrm F}\nonumber \\ = & W_{F} \log_2\Big(1 +  \gamma_{\text{UE}}^{\mathrm F}\Big) + W_{F} \log_2\Big(1 + \gamma_{S_2}^{\mathrm F}\Big)
\end{align}}with $\mathcal R_{A}^{\mathrm F}$ and $\mathcal R_{\text{ISL}}^{\mathrm F}$ being the achievable rates for access and backhaul, respectively, under FDD backhauling.

To maximize the throughput, optimal power control can be performed subject to (i) total satellite transmit power and (ii) minimum access rate constraints, as follows:
{\small \begin{align} \label{OP2}
	& \hspace{.1cm}\underset{\substack{P_{A},\ P_{I_1}}}{\text{maximize}} \hspace{0.3cm} W_{F} \log_2\Big(1 +  \frac{P_{A}\beta_{\text{UE1}}}{N_0W_F} \Big) + W_{F} \log_2\Big(1 + \frac{P_{I_{1}}\beta_I}{N_0W_F} \Big) \\ 
	&\hspace{.1cm}\text{subject to} \nonumber \\ 
	& \hspace{.1cm} P_{A} + P_{I_1} \le P_{S_1}, \label{2a} \\ & 
  \hspace{.1cm} W_{F} \log_2\Big(1 +  \frac{P_{A}\beta_{\text{UE1}}}{N_0W_F} \Big) \ge \overline{\mathcal R}, \label{2b} 
\end{align}}where $P_{S_1}$ is the total available transmit power at $S_1$, and $\overline{\mathcal R}$ is a minimum access rate threshold for the UE. Problem (\ref{OP2}) and constraints (\ref{2a}) and (\ref{2b}) are convex, and the optimal solution can be obtained using software tools such as the CVX. 
\section{Scenario 2: TDD for ISL} \label{sec: TDD}
For this scenario, the TDD mode is utilized for backhauling between $S_1$ and $S_2$, while FDD is adopted for the communication between $S_1$ and the UE for data access. It is worth highlighting that in this case, the transmission from $S_2$ to $S_1$ would likely cause interference to the UE due to the undesired side-lobes of the main beam that is intended for $S_1$. In particular, and unlike the previous scenario with FDD, when $S_2$ operates in TDD mode, the entire available bandwidth will be utilized for both transmission and reception (at different time slots). Similarly, the omni-directional transmission at the UE can also affect the received signal at $S_2$ when the latter operates in a TDD mode, although such interference at $S_2$ would be small in case of a single handheld UE transmission.
\subsection{Received Signals and SNRs}
 We start with the received signal at the UE, which can be expressed as:
\begin{equation}\small 
    y_{\text{UE}}^{\mathrm T} = \sqrt{P_{A}} h_{\text{UE}1} x_{\text{UE}} + \varpi \sqrt{P_{I_2}} h_{\text{UE}2} x_{I_1} + z_{\text{UE}},
\end{equation}where the superscript in $y_{\text{UE}}^{\mathrm T}$ indicates that the TDD mode is utilized for backhauling, $h_{\text{UE}2}$ is the channel between $S_2$ and the UE, $P_{I_2}$ is the transmit power from $S_2$,\footnote{In this work we assume that $P_{I_2}$ is equivalent to the total transmit power at $S_1$, i.e. $P_{I_2} = P_{S_1}$.} $x_{I_1}$ is the information symbol intended for $S_1$ from $S_2$ (via the ISL) with $\mathbb E\{|x_{I_1}|^2\}=1$, and the parameter $\varpi$ accounts for the fact that $S_2$ operates in the TDD mode and thus only transmits (to $S_1$) for half of the time. Therefore
\begin{equation} \small 
\varpi = 
    \begin{cases}
    0, \hspace{1cm} \text{when $S_2$ is in receiving mode from $S_1$} \\ 1, \hspace{1cm} \text{otherwise.}
    \end{cases}    
\end{equation}In addition, the received signal at $S_2$ (when operating in the receiving mode) is:
\begin{equation}\small 
y_{S_2}^{\mathrm T} = \sqrt{P_{I_{1}}} h_{I} x_{I_2} + \sqrt{P_{\text{UE}}} h_{\text{UE2}} x_{S_1} + z_{S_2},
\end{equation}where $P_{\text{UE}}$ is the UE uplink transmit power, and $x_{S_1}$ is the transmitted data symbol from the UE intended for $S_1$.

The SNR/SINR (signal-to-interference-plus-noise ratio) at the UE and $S_2$ are, respectively, given as follows:
\begin{subequations}
\begin{equation} \small 
    \gamma_{\text{UE}}^{\mathrm T} = \begin{cases}
        \frac{P_{A}\beta_{\text{UE1}}}{N_0W_F} , \hspace{2.53cm} \text{for $\varpi = 0$}, \\ \frac{P_{A}\beta_{\text{UE1}}}{P_{I_2}\beta_{\text{UE2}} \frac{W_F}{W_T} + N_0W_F}, \hspace{1cm} \text{for $\varpi = 1$}, 
    \end{cases}
\end{equation} 
\begin{equation}\small 
    \gamma_{S_2}^{\mathrm T} = \frac{P_{I_1}\beta_I}{P_{\text{UE}}\beta_{\text{UE2}} + N_0W_T},
\end{equation}    
\end{subequations}where $W_T$ is the total available bandwidth under the TDD mode.
\subsection{Achievable Rates and Power Control}
In this case, there are two different achievable rates at the UE, which can be expressed as:
\begin{subequations}
\begin{equation} \label{R_UE_TDD} \small 
    \mathcal R_{A_1}^{\mathrm T} = W_F \log_2\Big(1 + \frac{P_{A}\beta_{\text{UE1}}}{N_0W_F} \Big),
\end{equation}
\begin{equation} \small 
    \mathcal R_{A_2}^{\mathrm T} = W_F \log_2\Big(1 + \frac{P_{A}\beta_{\text{UE1}}}{P_{I_2}\beta_{\text{UE2}} \frac{W_F}{W_T} + N_0W_F}\Big),
\end{equation}
\end{subequations}
where $\mathcal R_{A_1}^{\mathrm T}$ reflects the achievable access rate under no interference from $S_2$ (i.e. $S_2$ is receiving from $S_1$), while $\mathcal R_{A_2}^{\mathrm T}$ is the achievable access rate at the UE when $S_1$ is transmitting to the UE but receiving from $S_2$.  

In addition, the backhaul achievable rate under TDD transmission is:
\begin{equation} \label{R_ISL_TDD}\small 
    \mathcal R_{\text{ISL}}^{\mathrm T} = \frac{1}{2} W_T \log_2\Big(1+ \frac{P_{I_1}\beta_I}{P_{\text{UE}}\beta_{\text{UE2}} + N_0W_T}\Big),
\end{equation}where the $\frac{1}{2}$ factor in (\ref{R_ISL_TDD}) is due to the TDD operation. 
\par The goal is to maximize the total throughput via power allocation at $S_1$. However, such power control is only required when $S_1$ is transmitting to both the UE and $S_2$, while for the second case where $S_1$ is receiving from $S_2$, both $S_1$ and $S_2$ can transmit with their maximum available powers to serve the UE and $S_1$, respectively.

Moreover, the optimization, which is performed to split the power during only one of the TDD time instances, should take into account the minimum average user rate as follows
{\small \begin{align} \label{OP3}
	& \hspace{.1cm}\underset{\substack{P_{A},\  P_{I_1}}}{\text{maximize}} \hspace{1cm} \Bigg\{W_F \log_2\Big(1 + \frac{P_{A}\beta_{\text{UE1}}}{N_0W_F} \Big) \nonumber \\ & \hspace{2.45cm} + \frac{1}{2} W_T \log_2\Big(1+ \frac{P_{I_1}\beta_I}{P_{\text{UE}}\beta_{\text{UE2}} + N_0W_T}\Big)\Bigg\} \\ 
	&\hspace{.1cm}\text{subject to} \nonumber \\ 
	& \hspace{.1cm} P_{A} + P_{I_1} \le P_{S_1}, \label{3a} \\ & 
  \hspace{.1cm} W_{F} \log_2\Big(1 +  \frac{P_{A}\beta_{\text{UE1}}}{W_{F} N_0} \Big) + \mathcal R_{A_2}^{\mathrm T} \ge 2\overline{\mathcal R}. \label{3b} 
\end{align}}Problem (\ref{OP3}) is convex and can be solved optimally using software tools. Once $\mathcal R_{A_1}^{\mathrm T}$ and $\mathcal R_{\text{ISL}}^{\mathrm T}$ are maximized via power control, the total network throughput can be expressed as: 
\begin{equation}\small 
\mathcal R^{\mathrm T} = \mathcal R_{\text{ISL}}^{\mathrm T} + \frac{1}{2}(\mathcal R_{A_1}^{\mathrm T} + \mathcal R_{A_2}^{\mathrm T}),
\end{equation}where the division over two is to find the average access rate with and without interference from $S_2$.
\begin{table}[t]
\centering
\caption{Simulation parameters.} \label{table3}
\begin{tabular}{|c|c|}
\hline 
\textbf{Parameter}             & \textbf{Value} \\ \hline
Total bandwidth ($W_T$)                & $40$ MHz         \\ \hline
FDD bandwidth ($W_F$)                & $0.5 W_T$         \\ \hline
Total satellite transmit power ($P_{S_1}$)              & $30-40$ dBm  \\ \hline
Earth Radius ($R_E$)              & $6371$ Km  \\ \hline
Noise PSD ($N_0$)    & $-174$ dBm/Hz  \\ \hline
Satellite antenna gain ($G_{S_1}$, $G_{S_2}$) & $32$ dBi         \\ \hline
UE antenna gain ($G_{\text{UE}}$)               & $0$ dBi          \\ \hline
Carrier frequency ($f_c$)         & $2$ GHz (FR1)          \\ \hline
LEO satellite altitude         & $\{600,\ 1200\}$ Km   \\ \hline
Minimum access rate ($\overline{\mathcal R}$)    & 10 Mbits/sec         \\ \hline
UE transmit power    & $20$ dBm         \\ \hline
\end{tabular}
\end{table}
\section{Numerical Evaluations}\label{sec: Results}
In this section, we present and discuss the numerical results for the adopted satellite IAB network. The different simulation parameters utilized in our work are shown in Table \ref{table3}. 

\begin{figure}[t!]
 \centering
       \includegraphics[width=7.9cm,height=6.1cm]{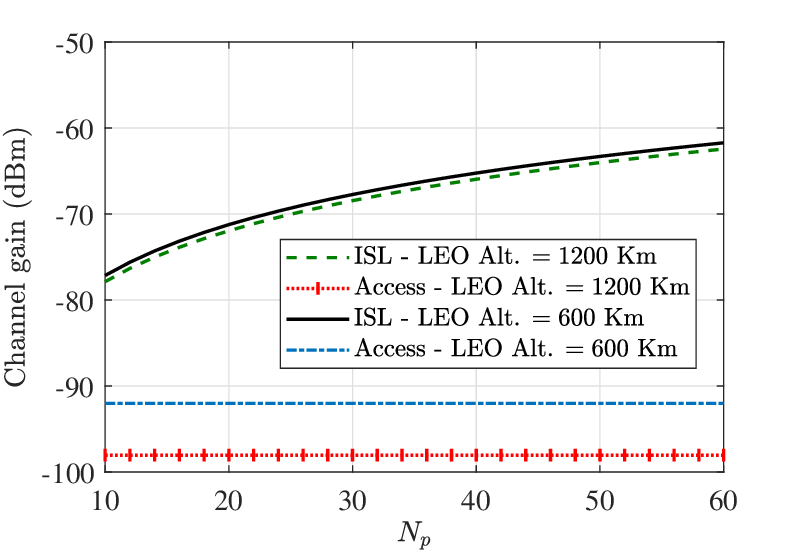} 
        \caption{The effect of number of satellites and LEO altitudes on channel conditions.}
        \label{Channel_gain}
\end{figure}

First, we show in Fig.~\ref{Channel_gain} the ISL and access channels gain by varying the number of LEO satellites in the considered orbital plane and satellites' altitude.\footnote{By applying the formula in (\ref{N_min}), the minimum number of LEO satellites to maintain an ISL connection was found to be $6$ for orbital planes with an altitude of $1200$ Km, and $8$ for orbital planes with an altitude of $600$ Km.} Clearly, the number of satellites in the orbital plane has a direct effect on the ISL link quality, as larger number of satellite nodes means shorter slant range between any two neighbouring satellites, and thus, better channel conditions. In this context, it is worth highlighting that the number of satellites per plane in the Starlink constellation (SpaceX) ranges between 20 and 58 satellites according to the Federal Communications Commission (FCC) report in \cite{SpaceX}. In addition, the results show that while the quality of the access link is highly affected by the LEO altitude, the effect on the quality of the ISL is marginal. The reason is that for LEO satellites, having twice the altitude does not correspond to having twice the slant range. In fact, in the considered scenario, the slant range at $1200$ Km of LEO altitude is only about $9\%$ larger than that at $600$ Km, assuming the same number of satellites at both altitudes. Furthermore, the ISL quality is shown to be much higher than that for the access link. This is a direct consequence of the fact that the handheld UE is assumed to have an omni-directional antenna with $0$ dBi antenna gain compared to $S_2$ which has $32$ dBi antenna gain.   

\begin{figure}[t!]
 \centering
       \includegraphics[width=7.9cm,height=6.1cm]{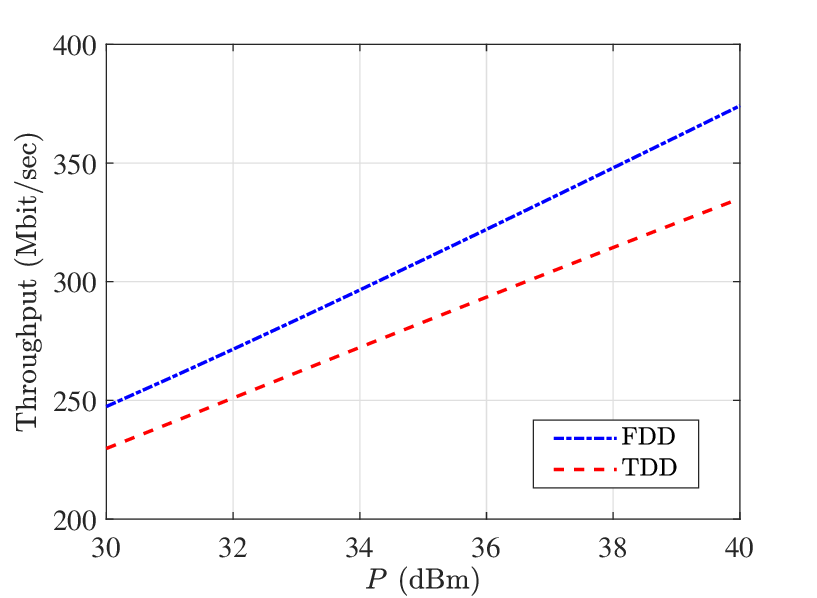} 
        \caption{Network throughput vs transmit power level at $S_1$ when $N_p = 30$ and LEO altitude is $600$ Km.}
        \label{Throughput_arch_2}
\end{figure}

Fig.~\ref{Throughput_arch_2} shows the network throughput as a function of the total transmit power at the satellite nodes, and for both TDD and FDD transmission modes under a minimum access rate of $\overline{\mathcal R} = 10$ Mbits/sec. The FDD demonstrates superior performance compared to its counterpart the TDD. Moreover, higher levels of satellite transmit powers lead to a bigger gap between the TDD and the FDD modes. The reason is that under the TDD transmission, one cannot escape the resultant interference between $S_2$ (which transmits with the same power level as $S_1$ of $P$ Watts) and the ground-based UE, and thus, higher transmit power levels mean larger interference, and hence, bigger gap between the two transmission modes.

\begin{figure}[t!]
 \centering
       \includegraphics[width=7.9cm,height=6.1cm]{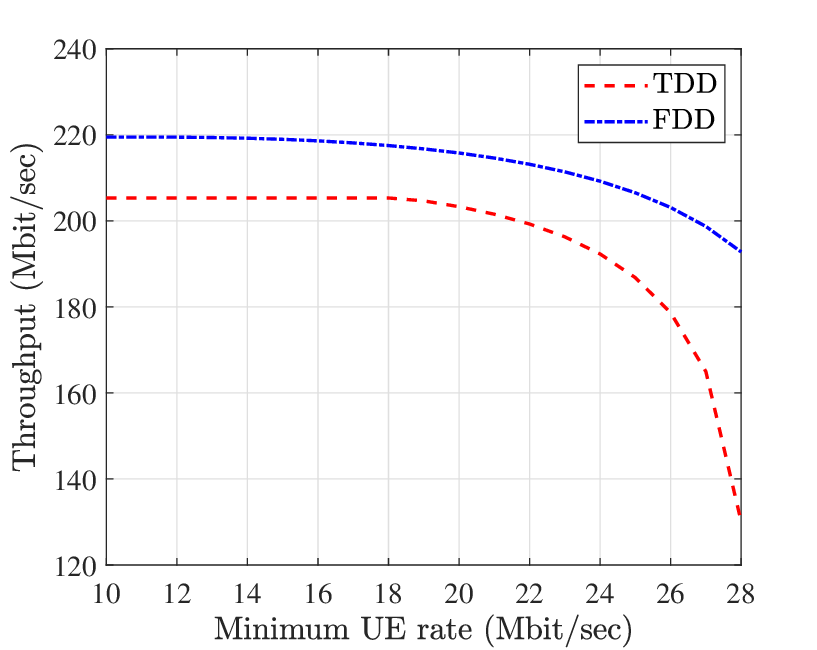} 
        \caption{{Network throughput vs minimum access} rate at UE when the total power at $S_1$ and $S_2$ is $30$ dBm, LEO altitude is $1200$ Km, and $N_p$ = 30.}
        \label{Minimum_UE_rate}
\end{figure}

Fig.~\ref{Minimum_UE_rate} demonstrates the network throughput (which comprises both access and backhaul achievable rates) under different access QoS requirements ($\overline{\mathcal R}$) at the UE. The results indicate a significant degradation in the total throughput is experienced as the minimum access rate increases, especially for the TDD scenario due to the interference between $S_2$ and the UE that makes it particularly challenging to guarantee a high QoS at the UE. For instance, in order for the UE to enjoy a minimum rate of 28 Mbits/sec, the total throughput will be degraded by 75 Mbits/sec compared to that which guarantees the UE with a QoS of only 10 Mbits/sec. 

\begin{figure}[t!]
 \centering
       \includegraphics[width=7.9cm,height=6.1cm]{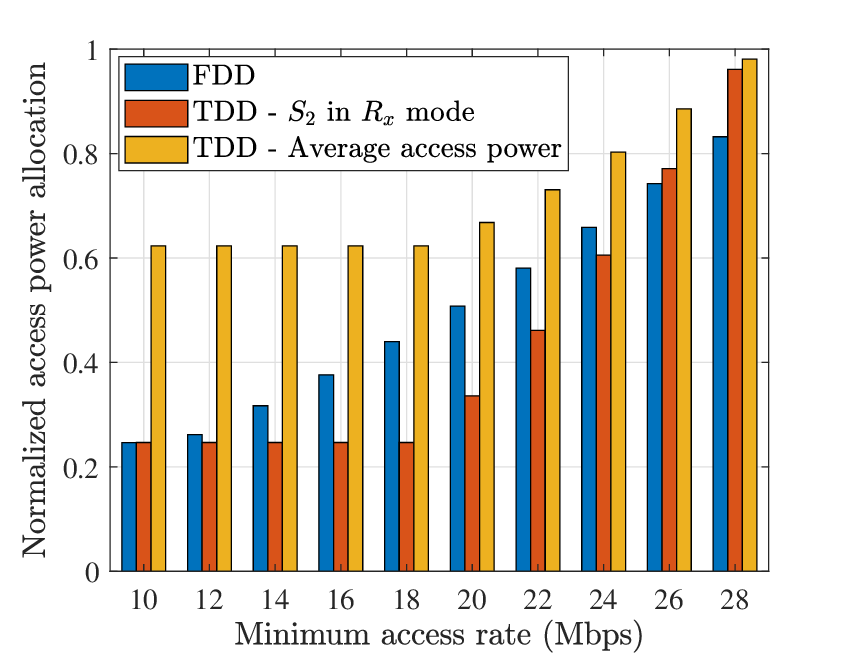} 
        \caption{Normalized access power allocation vs minimum access rate requirements under TDD and FDD for the ISL. Network parameters are identical to those in Fig.~\ref{Minimum_UE_rate}.}
        \label{fig:access_power}
\end{figure} 
The impact of higher QoS at the UE can be further seen in Fig.~\ref{fig:access_power}, which shows the distribution of the normalized allocated power for the access (i.e. allocated power for access divided by the total transmit power at $S_1$) as a function of the minimum required access rate, and for both TDD and FDD of ISL backhauling. The blue bar shows the normalized access power under the FDD scenario, while the red bar shows the amount of power for access during only one time instance of the TDD transmission, particularly the one where $S_1$ transmits to both the UE and $S_2$. Further, the yellow bar illustrates the average distribution of access power under the TDD scenario over both TDD time instances.\footnote{During the second time instant of the TDD scenario where $S_2$ is in transmission mode, it is assumed that $S_1$ deploys all of its available transmission power to serve the UE.} Focusing on the comparison between the red and blue bars where the total power at $S_1$ is split between the UE and $S_2$, $96\%$ of the total power is allocated for access to achieve a QoS of $28$ Mbit/sec under TDD, compared to $83\%$ under the FDD mode. This explains the sharp decrease in total network throughput under TDD for ISL when the UE QoS is relatively high. 

\begin{figure}[t!]
 \centering
       \includegraphics[width=7.9cm,height=6.1cm]{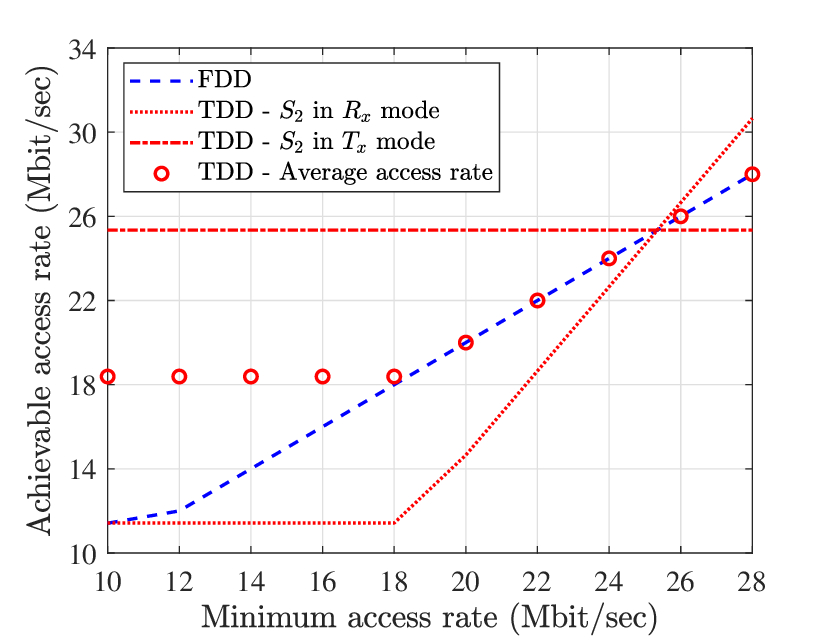} 
        \caption{Achievable access rate vs the minimum required access rate ($\overline{\mathcal R}$). Network parameters are identical to those in Fig.~\ref{Minimum_UE_rate}. }
        \label{fig:UE_rate}
\end{figure}

Finally, it is also observed from Fig.~\ref{Minimum_UE_rate} and Fig.~\ref{fig:access_power} that the throughput and power allocation do not experience any change under the TDD when the required QoS is below $20$ Mbit/sec. This can be explained in Fig.~\ref{fig:UE_rate}, where the average access rate (that also happens to maximize the network throughput) under the TDD transmission is above $18$ Mbit/sec. As a result, it is only when the minimum requirement is above this threshold that the network would have to compromise the total throughput by allocating higher power levels to the UE in order to achieve its required QoS. In contrast, under the FDD transmission, the optimal access rate that maximizes the total throughput is just above the $11$ Mbit/sec mark, therefore, any QoS requirement that is higher than this threshold will lead to compromising the total network throughput by modifying the power distribution between the access and backhaul links. 
\section{Concluding Remarks}\label{sec: Conclusions}
The performance of satellite IB-IAB with inter-satellite backhauling over the S-band frequency spectrum was investigated in this work. Following the 3GPP specifications, the FDD mode was adopted for the direct access between a handheld UE with omni-directional antenna and the LEO satellite, while both TDD and FDD transmission modes were investigated for backhauling through ISLs. Our results demonstrated the large superiority of FDD compared to TDD, and they showed that under a high QoS for the access link, the total network throughput will suffer dramatically if the backhauling was performed according to the TDD mode due to the interference between the access and backhaul links. 
\section*{Acknowledgment}
This work has been supported by the European Space Agency (ESA) funded activity Sat-IAB: Satellite and Integrated Access Backhaul - An Architectural Trade-Off (contract number 4000137968/22/UK/AL). The views of the authors of this paper do not necessarily reflect the views of ESA.
\section*{Appendix}
\begin{figure}[h]
    \centering
    \includegraphics[width=4.6cm,height=5.50cm]{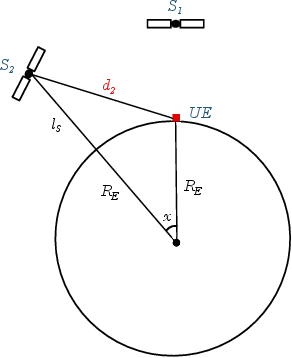}
    \caption{A geometric illustration of the considered system model.}
    \label{Illustration}
\end{figure}

From Fig. \ref{Illustration}, the distance between $S_2$ and the UE can be evaluated using the following trigonometric formula:
\begin{equation}\label{x}
\small 
    d_2 = \sqrt{(R_E + l_s)^2 + R_E^2 - 2(R_E+l_s)R_E\cos{x}},
\end{equation}where $R_E$ is the Earth's radius. Assume that there are $N_p$ evenly-distributed satellites within the considered orbital plane, $d_2$ can be obtained after substituting $x= 2\pi/N_p$ in~(\ref{x}).

\bibliographystyle{IEEEtran}
\bibliography{IAB_ISL}	
\end{document}